\documentclass[runningheads,a4paper]{llncs}

\usepackage{amssymb}
\usepackage{graphicx}
\usepackage{algpseudocode}
\usepackage{enumerate}
\usepackage{amsmath}
\usepackage{xspace}
\usepackage{alltt}
\usepackage{authblk}
 \usepackage{color, colortbl}
\usepackage{url}
\urldef{\mailsa}\path|{alfred.hofmann, ursula.barth, ingrid.haas, frank.holzwarth,|
\urldef{\mailsb}\path|anna.kramer, leonie.kunz, christine.reiss, nicole.sator,|
\urldef{\mailsc}\path|erika.siebert-cole, peter.strasser, lncs}@springer.com|

\newcommand{\keywords}[1]{\par\addvspace\baselineskip\noindent\keywordname\enspace\ignorespaces{#1}}
\newcommand{\remove}[1]{}
\newcommand{\VIP}{\textsc{Vip}}
\usepackage{color}
\newcommand{\commentout}[1]{%
}
\newcommand{\secref}[1]{Section~\ref{#1}}

\newcommand{\figref}[1]{Figure~\ref{#1}}
\newcommand{\eqrefx}[1]{Eq.~\ref{#1}}
\newcommand{\tabref}[1]{Table~\ref{#1}}
\newcommand{\modelch}[1]{\textsc{{#1}}\noindent\normalsize}
\definecolor{LightCyan}{rgb}{0.88,1,1}
\definecolor{Gray}{gray}{0.7}

\urldef{\mailsa}\path{jeonhyuk@usc.edu,lerman@isi.edu}

\begin{document}

\mainmatter

\title{VIP: Incorporating Human Cognitive Biases in a Probabilistic Model of Retweeting}

\titlerunning{VIP: Incorporating Human Cognitive Biases in a Probabilistic Model}
\author{Jeon-Hyung Kang \and Kristina Lerman}
\institute{USC Information Sciences Institute, 4676 Admiralty Way, Marina del Rey, CA
\email{jeonhyuk@usc.edu,lerman@isi.edu}}

\maketitle
\begin{abstract}

Information spread in social media depends on a number of factors, including how the site displays information, how users navigate it to find items of interest, users' tastes, and the `virality' of information, i.e., its propensity to be adopted, or retweeted, upon exposure. Probabilistic models can learn users' tastes from the history of their item adoptions and recommend new items to users. However, current models ignore cognitive biases that are known to affect behavior. Specifically, people pay more attention to items at the top of a list than those in lower positions. As a consequence, items near the top of a user's social media stream have higher visibility, and are more likely to be seen and adopted, than those appearing below.
Another bias is due to the item's fitness: some items have a high propensity to spread upon exposure regardless of the interests of adopting users.
We propose a probabilistic model that incorporates human cognitive biases and personal relevance in the generative model of information spread. We use the model to predict how messages containing URLs spread on Twitter. Our work shows that models of user behavior that account for cognitive factors can better describe and predict user behavior in social media.

\end{abstract}

\section{Introduction}
Online social networks can dramatically amplify the spread of information by allowing users to forward information to their followers, and those to their own followers, and so on. Predicting how people will respond to information is of immense practical and commercial interest. Prediction can guide the design of more effective marketing and public awareness campaigns, for example, those announcing the locations of clinics dispensing the flu vaccine.
Researchers believe that information spread in social media  is a complex process that depends on the nature of information~\cite{Romero11www}, the structure of the network~\cite{bakshy2012role,weng2013virality}, the strength of social influences~\cite{Bakshy11,Romero11}, as well as user interests and topic preferences~\cite{Kang13sbp,Kang13aaai}.
These factors are thought to render information diffusion in social media unpredictable~\cite{Bakshy11,Goel12}, although researchers have identified some features that weakly correlate with the size of information cascades~\cite{Cheng14}.
On the other hand, more progress has been made addressing information spread in social media as a social recommendation problem. In this case, probabilistic models are used to learn users' topic preferences from the history of their item adoptions and predict what items in their social media stream users will adopt ~\cite{ma2008sorec,WangB11,Kang13aaai}.

Existing models of social recommendation largely ignore cognitive factors of user behavior. One such aspect is \emph{position bias}. Due to this cognitive bias, the amount of attention an item receives strongly depends on its position on a screen or within a list of  items. Position bias is known to affect the answers people select in response to multiple-choice questions~\cite{Payne51,Blunch84}, where on the screen they look~\cite{buscher2009you,Counts11}, and the links on a web page they choose to follow~\cite{Craswell08,Huberman:1998eq}. Also as a consequence of position bias, items near the top of a user's social media stream are more salient, and therefore, more likely to be viewed, than items in lower positions~\cite{Lerman14plosone}.

To distinguish position-based salience from other psychological effects, we refer to it as an item's \emph{visibility}. After viewing the item, the user may decide to adopt it. She adopts information either because it is personally relevant to her or because it is generally interesting. To handle the former case, the model must include a hidden topic space which can be used to compute the {relevance} of items to users. The user may also adopt an item that is not strictly relevant, but interesting nonetheless. Such items are usually viral memes, such as breaking news, that have a high {fitness}, i.e., propensity to spread upon exposure regardless of the interests of adopting users.

In \secref{sec:lactr}, we introduce a conceptually simple model of information spread  that captures the factors important to information spread: item's \emph{visibility}, \emph{fitness}, and its \emph{personal relevance} to user's interests.
An item's visibility depends on its position in the user's social media stream. However, since position data is often not directly available, we estimate visibility from user's information load. This quantity measures the number of items a user has to inspect before finding a specific item to adopt, and it is given by the number of new messages arriving in the user's stream and the frequency the user visits the stream. The greater the number of new messages in the stream --- either because the user follows more people or because she rarely visits her stream --- the less visible any particular item is.
Accounting for visibility allows us to learn a better model of user's interests from the history of her item adoptions. When the user does not adopt an item, the model allows us to discriminate between lack of interest and failure to see the item.
While this simple model ignores some of the nuances of information spread, it has very high predictive power.
 In \secref{sec:evaluation}, we evaluate the proposed  model on a social recommendation task using Twitter. We study the impact of visibility, item fitness, and personal relevance on information diffusion in Twitter in aggregate and through illustrative individual examples.
Our study demonstrates that models of user behavior that account for cognitive biases can better describe and predict user behavior in social media, and that information spread is more predictable than previously thought.

\section{The VIP Model}
\label{sec:lactr}
We describe {\VIP}, a model that captures the three basic ingredients of information spread in social media: item's fitness and its visibility and personal relevance to the user. {\VIP} is based on social recommendation models, whose goal is to recommend only the relevant items to users~\cite{ma2008sorec,WangB11,Kang13aaai,Kang13sbp}. In social recommendation, each user is assigned a vector of topics, which serve as her interest profile, and each item also has some topics. Once these hidden vectors are learned from the history of user item adoptions, it is possible to calculate an item's \emph{personal relevance} to the user.

Social media users adopt items even if they had not earlier demonstrated a sustained interest in their topics. This is often the case with viral, general-interest items, such as breaking news or celebrity gossip. We use the term \emph{fitness} (or `virality') to describe an item's propensity to be adopted upon exposure.

The key innovation of {\VIP} is to introduce \emph{visibility} into the generative model of item adoption. Visibility conceptually simplifies the mechanisms of information spread and explains away some of the complexity associated with it, for example, the network effects observed by \cite{Bakshy11,Romero11,Romero11www}. Visibility explicitly takes into account the process of information discovery in social media. Online social networks are directed, with users following the activities of their friends. A user's message stream contains a list of items her friends adopted or ``recommended'' to her, chronologically ordered by their adoption time, with the most recent item at the top of the stream.
We consider a user to be \emph{exposed} as soon as the item enters her stream; however, exposure does not guarantee that the user will actually view the item. The probability of viewing --- \emph{visibility} --- depends on the item's position in the user's stream~\cite{Hodas12socialcom}. Due to a cognitive bias known as position bias~\cite{Payne51}, a user is more likely to attend to items near the top of the screen than those deeper in the stream~\cite{buscher2009you}.
Below we discuss a method to quantitatively account for this visibility.

\begin{figure}[t]
\begin{center}
\includegraphics[width=0.4\linewidth]{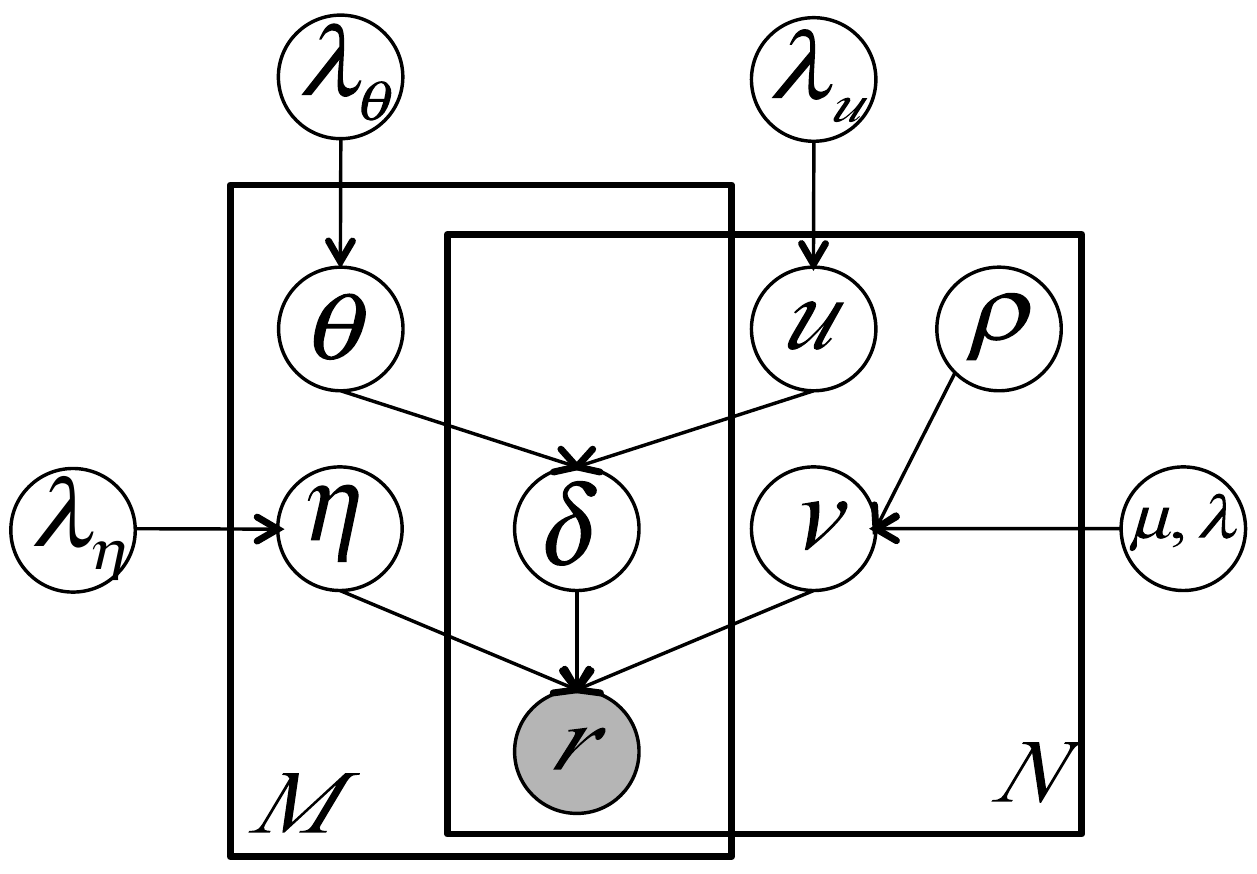}
\end{center}
\caption{The {\VIP} model  with user topic ($u$) and item topic ($\theta$) profiles, personal relevance of an item to user ($\delta$), visibility to user ($v$), item fitness ($\eta$), expected number of new posts user received ($\rho$) and item adoption ($r$).  $N$ is the number of users and ${M}$ is the number of items. }
\label{fig:modelDiagram}
\end{figure}

Figure~\ref{fig:modelDiagram} graphically represents the  {\VIP} model.
It considers a user $i$ with a user-topic vector $u$, and an item $j$ with an item-topic vector $\theta$. {\VIP} generates an adoption of an item $j$ by user $i$ as follows:
\begin{equation}
\begin{aligned}
& r_{ij} \sim {\mathcal{N} \left(   v_i g_r( \delta_{ij} + \eta_j), c_{ij}^{-1}  \right) } \\
 \end{aligned}
\end{equation}
where $\eta_j  \sim   \mathcal{N}  (0, \lambda_\eta^{-1})$, is the fitness (or interestingness) of item $j$, which represents the probability of adoption given the user viewed it~\cite{Hogg12epj,wang2013quantifying}. The precision parameter  $c_{ij}$  serves as confidence for adoption $r_{ij}$. $v_i$ represents the visibility of item to user and $\delta_{ij}$ represents user $i$'s interest in item $j$.
We define $g_r$ as a linear function for simplicity. One of the key properties of {\VIP} lies in how a user adopts items that have the same visibility. We assume that user adopts either items that are relevant to her ($\delta_{ij}$) or interesting in general ($\eta_j$).

Users discover items by browsing through their message stream. As argued above, the position of an item in the stream determines its visibility, the likelihood to be viewed. However, an item's exact position is often not known. Instead, we estimate its average visibility from the available data. This quantity depends on user's information load~\cite{Rodriguez14}, i.e., the flow of messages to the user's stream, and the frequency the user visits the site. The greater the number of new messages user receives between visits to the site, the less likely the user is to view any specific item.
Following \cite{Hogg13socialcom}, we estimate visibility of an item  to user $i$ as:
\begin{equation}
\begin{aligned}
& v_{i} \sim \sum_{L} \left( \mathbf{G}(1/(1+\rho_i),L) (1-\mathbf{IG}(\mu, \lambda, L)) \right) \\
 \end{aligned}
  \label{eq:visibile}
\end{equation}
The first factor gives the probability that $L$ newer messages have accumulated in user $i$'s stream since the arrival of a given item. The accumulation of items is a competition between the rates friends post new messages to the user's stream and the rate the user visits the stream to read the messages. The ratio $\rho_i$ of  these rates gives the expected number of new messages in a user $i$'s stream since item $j$'s arrival. Taking friends activity and user activity each to be a Poisson process, the competition gives rise to a geometric distribution with success probability $p=1/(1+\rho_i)$: $\mathbf{G} = (1-p)^L p$.
We will revisit how we estimate $\rho_i$ in the section below. The second factor of \eqrefx{eq:visibile} gives the probability that user $i$ will navigate to at least $L+1$'st position in her stream to view the item. This is given by  the upper cumulative distribution of an inverse gaussian $\mathbf{IG}$ with mean $\mu$ and shape parameter $\lambda$ and variance $\mu^3/\lambda$:
\begin{equation}
\begin{aligned}
& \exp \left(\frac{-\lambda  (L-\mu )^2}{2\mu ^2L} \right) \left[\frac{\lambda }{2\pi L^3} \right]^{(1/2) }.
 \end{aligned}
\end{equation}
This distribution has been used to describe the ``law of surfing''~\cite{Huberman:1998eq}, and it represents the probability the user will view $L$ items on a web page before stopping. Therefore, the cumulative distribution of $\mathbf{IG}$ gives the probability the user will view at least $L$ items, hence, navigating to $L+1$'st position in her stream.

We calculate personal relevance of the item $j$ to user $i$ as:
\begin{equation}
\begin{aligned}
& \delta_{ij} \sim g_\delta({u_i^T \theta_{j}})\\
 \end{aligned}
\end{equation}
where symbol $T$ refers to the transpose operation, $u_i$ represents topic profile of user $i$, $\theta_j$ represents topic profile of item $j$ and $g_\delta$ is linear function for simplicity. We represent topic profiles of users and items in a shared low-dimensional space as follows.
\begin{equation}
\begin{aligned}
& u_{i} \sim \mathcal{N}(0,  \lambda_{u}^{-1}I_{K})\\
& \theta_{j} \sim \mathcal{N}(0,  \lambda_{\theta}^{-1}I_{K})\\
 \end{aligned}
\end{equation}
where $K$ is the number of topics.
Note that if we only use personal relevance ($\delta$) and ignore visibility and fitness, {\VIP} model reduces to probabilistic matrix factorization (PMF) model~\cite{salakhutdinov2008probabilistic} that learns latent topics from user--item adoptions.

The generative process for item adoption through a social stream can be formalized as follows:
\begin{tabbing}
For \=each user $i$ \\
\>  Generate \= $u_{i} \sim \mathcal{N}(0,\lambda_{u}^{-1}I_{K}$)\\
\>  Generate \= $v_{i} \sim \sum_{l} \left( \mathbf{G}(1/(1+\rho_i),l) (1-\mathbf{IG}(\mu, \lambda,l)) \right)$ \\
For \=each item j \\
\>  Generate \= $\theta_{j} \sim \mathcal{N}(0,\lambda_{\theta}^{-1}I_{K}$)\\
\>  Generate \= $\eta_{j} \sim \mathcal{N}(0,\lambda_{\eta}^{-1}$)\\
 For \=each  user $i$\\
\> For each recommended item $j$ from friends\\
\>\;\;\;\;\;\;Generate the adoption $r_{ij} \sim {\mathcal{N} \left(   v_i g_r( \delta_{ij} + \eta_j), c_{ij}^{-1}  \right) } $
\end{tabbing}
Here $\delta_{ij}=u_i^T\theta_j$, $\lambda_u = \sigma_r^2/\sigma_u^2$, $\lambda_\theta$ = $\sigma_r^2/\sigma_\theta^2$, and $\lambda_\eta = \sigma_r^2/\sigma_\eta^2$.
Lack of adoption by user $i$ of item $j$ ($r_{ij}=0$) can be interpreted in two ways: either user saw the item but did not like it, or user did not see the item but may have liked it had she seen it.
While other models partly account for lack of knowledge about non-adoptions using smoothing~\cite{WangB11},  we properly model visibility of items to users.
We set $c_{ij}$ to a high value $a^r$ when $r_{ij}=1$ and a low value $b^r$ for  items recommended by friends and $c^r$ for the rest when $r_{ij}=0$ ($a^r>b^r>c^r>0$). In this paper, we use the confidence parameter values, $a^r=1.0$, $b^r=0.03$ and $c^r=0.01$, for $c_{ij}$.

\subsection{Learning Parameters}

To learn model parameters, we follow the approaches of \cite{WangB11,Kang13aaai} and develop coordinate ascent, an EM-style algorithm, to iteratively optimize the variables \{$u_i$,  $\theta_j$, $\eta_{j}$\} and calculate the maximum a posteriori estimates. MAP estimation is equivalent to maximizing the complete log likelihood ($\ell$) of $U$, $V$, $\theta$, $\eta$ and ${R}$ given $\lambda_u$, $\lambda_\theta$, $\lambda_\eta$, $\mu$, $\lambda$ and $\rho$.
 \begin{equation}
\begin{aligned}
\ell =
& -\frac{\lambda_u}{2} \sum_i^N u_i^T u_i
+\sum_i^N \log\left(\sum_l^L (1/\rho_i+1)(\rho_i/\rho_i+1)^l (1-\mathbf{IG}(\mu,\lambda,l)  \right)\\
& -\frac{\lambda_\theta}{2} \sum_j^M { \theta_j^T  \theta_j}-\frac{\lambda_\eta}{2} \sum_j^M {\eta_{j}}^T {\eta_{j}} -\frac{c_{ij}}{2}  \sum_i^N\sum_j^{M}  {(r_{ij} - v_i({ \delta_{ij}+\eta_j}))^2 } 
 \end{aligned}
\end{equation}
Given a current estimate, we take the gradient of $\ell$ with respect to $u_i$,  $\theta_j$, and $\eta_{j}$ and set it to zero. The update equations are:
\begin{equation*}
\begin{aligned}
&u_i \leftarrow {\left( \lambda_u I_k +   \Theta v_i C_{i}  v_i \Theta^T \right)}^{-1}   \Theta C_{i} \left( v_i R_i - v_i  \eta  v_i \right) \\
&\theta_j \leftarrow  {\left( \lambda_\theta I_k +   U V C_{j}  V U^T \right)}^{-1}  U C_{j} \left( V R_i -  \eta_j V V I_N \right) \\
&\eta_j \leftarrow {\left( \lambda_\eta +  v^T  C_j v \right)}^{-1}
v^T C_j  \left( R_j  - V U^T \Theta_j  \right)\\
 \end{aligned}
\end{equation*}
where $C_j$ is a diagonal matrix of confidence parameters $c_{ij}$. Item visibility to user $i$, $v_i$, is represented as a diagonal matrix $V$ or in vector format as $v$. We define $\Theta$ as $K \times M$ matrix, $U$ as $K \times N$ matrix and $R_j$ as vector with $r_{ij}$ values for all pairs of users $i$ for the given item $j$.
\subsection{Prediction}
After  parameters are learned, {\VIP} can be used to predict item adoptions by a user.
For user-item adoption prediction, user $i$'s adoption of item $j$ retweeted by a friend is obtained by point estimation with optimal variables \{$\theta^{*}$, $u^{*}$, $v^{*}$, $\eta^{*}$\}:
\begin{equation}
\begin{aligned}
\mathbb{E}[r_{ij}|\mathcal{D}] \approx&
     \mathbb{E}{[v_{i}|\mathcal{D}]}^T  (\mathbb{E}[\delta_{ij}|\mathcal{D}] +\mathbb{E}[\eta_{j}|\mathcal{D}] )\\
r_{ij}^{*} \approx& {v_{i}^{*}} ( {u_{i}^{*}}^T{\theta_{j}^{*}}+{\eta_{j}^{*}})
 \end{aligned}
\end{equation}
where $\mathcal{D}$ is the training data. The adoption probability is decided by user visibility $v^{*}_{i}$, user topic profile  $u^{*}_i$, item topic profile $v^{*}_j$, and item fitness  $\eta^{*}_j$.

\section{Evaluation}
\label{sec:evaluation}
In this section we demonstrate the utility of the {\VIP} model by applying model to data from the social media Twitter and evaluating its performance on the prediction tasks.
We collected tweets containing a \emph{URL} to monitor information spread over the social network from Nov 2011 to Jul 2012. We start by monitoring potential seed \emph{URL}s from streaming APIs and collected the entire history using the Twitter REST APIs to reconstruct their sharing history. This yielded 12.5M tweets with 9.5M users.

\subsection{Model Selection}
\label{sec:ms}
First, we study how parameters of  {\VIP} affect the overall performance of user-item adoption prediction using $recall@3$. We use the same ``law of surfing'' parameters, $\mu=14.0$ and $\lambda=14.0$, as \cite{Hogg13socialcom,Hogg12epj} did in their study of Twitter and another social media site. The expected number of new posts including a URL user $i$ received, $\rho_i$, is computed by $rate^{(url\ posts\ received)}_i/rate^{(visits)}_i$. The rate $rate^{(posts\ received)}_i$ is proportional to the  number of friends ($N_{frd(i)}$) $i$ follows and their average posting frequency~\cite{Hogg13socialcom}. To estimate posting frequency of all users, we have to track all their behaviors. Instead of tracking all users, we estimate it using the typical URL posting rates of users from our data: $rate^{(posts\ received)}_i=1.4*N_{frd(i)}$. User $i$ visits Twitter at a rate $rate^{(visits)}_i$. This number is not available; however, we expect it to be proportional to the number we do observe: the number of posts of user $i$ ($N_{posts(i)}$).  \cite{Hogg13socialcom} estimated that average number of visits per post was 38 for Twitter users. Also, since around 20\% of tweets include a URL~\cite{chaudhry2012trends}, the posting rate of user $i$ becomes $rate^{(visits)}_i = 7.6*N_{posts(i)}$.

For the PMF model, we vary the parameters $K\in $\{10, ... , 200\}, $\lambda_u$ and $\lambda_\theta \in $$\{10^{-4},..., 10^4\}$ by using grid search on validation recommendations. Throughout this paper, we set parameters $K=30$, $\lambda_u  =10^{-3}$, $\lambda_\theta  = 10^{-3}$ both for PMF and {\VIP} that performed the best for PMF. For the fitness parameter, we vary $\lambda_\eta \in \{10^{-4},..., 10^{4}\}$, while we fix $\lambda_\theta=10^{-3}$ and $\lambda_u=10^{-3}$ and set $\lambda_\eta=10^{4}$.
\subsection{ User--Item Adoption Prediction}
In the prediction task, we sort the items by $r_{ij}$, the probability of adoption by user $i$, and calculate the fraction of the ${X}$  top-ranked items  that the user actually adopted.
A user may not adopt an item either because she did not see it or because she does not like it. This makes it difficult to use precision to evaluate prediction results. Instead, we use $recall@{X}$ (=N(items in top \emph{X} user adopted)/N(items user adopted)) to measure model's performance on the prediction task.
To summarize performance of the prediction algorithm, we average recall values over all users.

We divide each user's adopted items into five folds and construct the training set and the test set. We use five-fold cross validation and compare performance of {\VIP} to three baseline models: \modelch{Random}, \modelch{Fitness} and \modelch{Relevance}. The \modelch{Random} baseline chooses items at random from among the items in user $i$'s stream, i.e., items adopted by $i$'s friends. The baseline \modelch{Fitness} uses item fitness values ($\eta$) learned by {\VIP} to recommend $X$ highest fitness items. The baseline \modelch{Relevance} bases its recommendations on user-topic and item-topic vectors learned by PMF to recommend $X$ most relevant items.

\begin{figure*}[tbh]
\begin{center}
\begin{tabular}{ccc}
\includegraphics[width=0.33\linewidth]{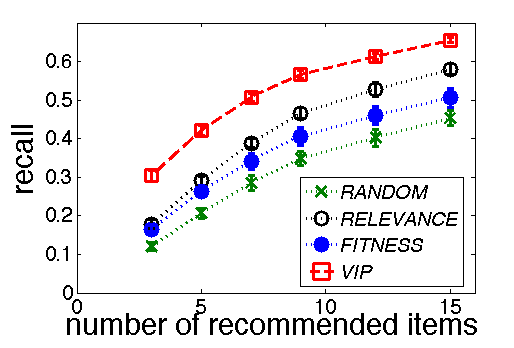}&
\includegraphics[width=0.33\linewidth]{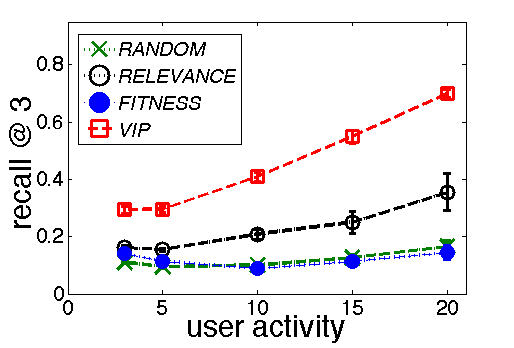}&
\includegraphics[width=0.33\linewidth]{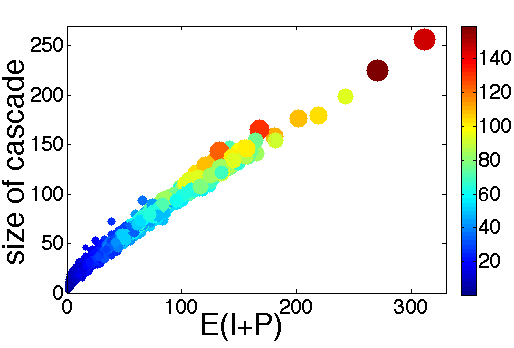}
\\(a)&(b)&(c)
\end{tabular}
\end{center}
\caption{Error bars are shown indicating standard deviation with the upper bar and the lower bar. (a)Recall of user--item adoption prediction with different numbers of recommended items. The number of topics was fixed at 30. (b)Average $recall@{3}$ of user--item adoption prediction for different activity levels of users with 30 topics. (c)Cascade size vs expected values of \emph{item fitness} plus \emph{personal relevance} E(I+P) for all adopters. The size and color of each circle represents the expected value of that item's \emph{visibility}. }
\label{fig:compiledfigs}
\end{figure*}

\figref{fig:compiledfigs} (a) shows the models' overall performance on the user--item adoption prediction task when we vary $X$, the number of recommendations made by each model. Note that a better model should provide higher $recall@{X}$ for different $X$.
{\VIP} outperforms all baselines, but the improvement is especially dramatic when the number of recommended items is small: $recall@{3}$ was 0.30, 0.17, 0.16, and 0.12 for {\VIP}, \modelch{Relevance}, \modelch{Fitness}, and \modelch{Random} models respectively. Note that as the number of recommendation (${X}$) increases, the recall of all models improves, however, at the expense of precision.

\figref{fig:compiledfigs} (b) shows how prediction performance on the user--item recommendation task varies with user activity level, that is how the number of items adopted by the user in the training set affects $recall@3$ on the test set. The performance of the \modelch{Random} baseline, which recommends three randomly chosen items from the user's stream, does not vary with user activity level, as expected.  Similarly, \modelch{Fitness} baseline does not vary significantly with activity, since it depends only on the propensity of the item to spread.  Both {\VIP} and \modelch{Relevance} improve with increasing user activity as they can learn better user--topic profiles with more training data. Note that for low activity users, whose interests are not well-known, recommending items based on personal \modelch{Relevance} performs about the same as picking items based on their fitness, but as more can be learned about user's preferences, \modelch{Relevance} outperforms picking items based on their fitness or picking them randomly from the user's stream.
{\VIP} handily outperforms baselines over all user activity levels. This shows that accounting for visibility dramatically improves predictability of user item adoptions in social media compared to using personal relevance or item fitness alone.

\section{Visibility \lowercase{vs} Item Fitness \lowercase{vs} Personal Relevance}
\label{sec:analysis}

We analyze URL cascades on Twitter by examining how the three factors learned by the {\VIP} model contribute to their success.
 Depending on the characteristics of the community that the URL has reached, fitness can vary.
In our data set, URLs that have been retweeted within a community sharing a specific hobby or interest, tend to have high fitness values. Since members share common topic preferences, items received relatively high adoption rates per exposure, which also often translates into quick adoption.
High fitness means high adoption rates per view
with statistically significant 0.85 correlation with cascade size. However, 4\% of the URLs have fitness values that are negatively correlated with cascade size and 40\% of the URLs show no correlation between fitness and cascade size. Apparently fitness by itself cannot explain the spread of information, and other factors, such as {visibility} and {personal relevance} also have to be considered.

\begin{table*}
  \centering
\caption{Cascade size, expected values, descriptions on Youtube video URLs}
\scalebox{0.95}{
\scriptsize{\begin{tabular}{|c|c|c|c|c|c|c|} 
  \hline
\cellcolor{Gray}{\scriptsize{Descriptions}}	&	\cellcolor{Gray}{\scriptsize{Cascade Size}	}&	\cellcolor{Gray}{\scriptsize{E(V)}}	&	\cellcolor{Gray}{\scriptsize{E(I)}}	 &	\cellcolor{Gray}{\scriptsize{E(P)}}	\\
\hline
\emph{	Strongbow surfers Neon Night Surfing on Bondi Beach	}	&	84	&	49	&	-0.04	&	85.1	\\
\hline
\emph{	Jay-Z Music Video	}	&	141	&	50.5	&	-0.04	&	130.7	\\
\hline
\emph{	Parallels for Mac for Chrome OS and Windows 8	}	&	68	&	52.3	&	-0.05	&	71.6	\\
\hline
\emph{	Bahraini Activist Nabil Rajab	}	&	116	&	62.1	&	-0.03	&	127.6	\\
\hline
\emph{	Ellen's Swaggin' Wagon -- a marriage Proposal}	&	87	&	65.6	&	 -0.13	&	80.2	\\
\hline
\emph{	UNICEF - Making headway toward an AIDS-free generation	}	&	102	&	71.7	&	-0.11	&	100.6	 \\
\hline
\emph{	Whitney Houston Video	}	&	120	&	73.5	&	-0.13	&	127.9	\\
\hline
\emph{	Paul McCartney's message from Moscow	}	&	109	&	87	&	-0.22	&	102.2	\\
\hline
\emph{	Ian Somerhalder Foundation	}	&	143	&	98.6	&	-0.24	&	150.8	\\
\hline
\end{tabular}
}}
\label{tbl:youtube}
\end{table*}
We separate the effect of item quality from its visibility to the user. We define quality very loosely as the combined effect of its fitness and relevance to adopters, and measure it by the expected value of these variables. This definition aims to make quality specific to the item itself, and separate from the details of how users may discover it.  \figref{fig:compiledfigs} (c) shows how the size of cascades in our data set depends on item quality and visibility. Each circle represents a URL, with its color encoding the expected visibility of the URL. Not surprisingly, higher quality URLs have larger cascades. More interestingly, some of the variance of cascade size can be explained by visibility: for URLs of similar quality, the more visible URLs spread more widely. In other words, for items cascading through a network where users have similar topic preferences, the total size of the cascade is decided by their {visibility}.

Next we illustrate the contributions of the three factors using specific case studies. There were 205 URLs to Youtube videos in our data set, with examples shown in \tabref{tbl:youtube}. Two of the most popular URLs in our data set were \emph{``Jay-Z Music Video''} and \emph{``Ian Somerhalder Foundation ''}, which were both adopted more than 140 times through friends' recommendation.
The fitness of the \emph{``Jay-Z Music Video''} is six times higher than that of \emph{``Ian Somerhalder Foundation''}, but has half the expected {visibility}. Therefore, the high fitness value of \emph{``Jay-Z Music Video''} makes up for the relatively low {visibility} and reaches a similar number of adoptions as {``Ian Somerhalder Foundation''}.

\section{Conclusion}

In this paper, we proposed {\VIP}, a model that captures the mechanisms of information spread in social media. {\VIP} can recommend items  to users  based on how easily users find an item in their stream, how well the item aligns with their interests, and the item's propensity to be adopted upon exposure. Prediction is surprisingly accurate, considering the crude estimates of visibility. Knowing visibility more accurately will further improve prediction performance. We plan to extend our model to take into account descriptions of items. We will further study the role of network structure and the relationship between visibility, item fitness, and personal relevance on information sharing.

\section*{Acknowledgments}
This  work was supported in part by AFOSR (contract FA9550-10-1-0569), by NSF (grant CIF-1217605)  and by DARPA (contract W911NF-12-1-0034).

\bibliographystyle{abbrv}
\bibliography{sigproc}

\end{document}